\documentclass[conference]{IEEEtran}
\IEEEoverridecommandlockouts
\usepackage{cite}
\usepackage{amsmath,amssymb,amsfonts}
\usepackage{graphicx}
\usepackage{textcomp}
\usepackage{xcolor}

\usepackage{graphicx}
\usepackage{dcolumn}
\usepackage{bm}
\usepackage{xcolor}
\usepackage{qcircuit}
\usepackage{braket}
\usepackage{float}
\usepackage{algorithm}
\usepackage{algorithmicx}
\usepackage{algpseudocode}
\usepackage{mathtools}
\usepackage{subcaption} 
\usepackage{hyperref}
\usepackage{graphicx}

\usepackage{amsthm}

\def\BibTeX{{\rm B\kern-.05em{\sc i\kern-.025em b}\kern-.08em
    T\kern-.1667em\lower.7ex\hbox{E}\kern-.125emX}}

\theoremstyle{definition}
\newtheorem{definition}{Definition}[section]

\renewcommand{\figureautorefname}{Figure~\negthinspace}
\renewcommand{\equationautorefname}{Equation~\negthinspace}

\renewcommand{\sectionautorefname}{Section~\negthinspace}

\begin{document}

\title{Quantum Machine Learning Architecture Search via Deep Reinforcement Learning
\thanks{This work was supported by the U.S. DOE, Office of Science, Office of High Energy Physics under award DE-SC-0012704.  This research used resources of the NERSC, under Contract No.DE-AC02-05CH11231 using NERSC award HEP-ERCAP0023403.}
}

\author{
Xin Dai$^1$, Tzu-Chieh Wei$^2$, Shinjae Yoo$^1$, Samuel Yen-Chi Chen$^1$\\
$^1$Computational Science Initiative, Brookhaven National Laboratory $\quad$ \\ $^2$C.N. Yang Institute for Theoretical Physics and Department of Physics and Astronomy, Stony Brook University $\quad$ \\
\small \texttt{\{xdai, sjyoo\}@bnl.gov, tzu-chieh.wei@stonybrook.edu, ycchen1989@ieee.org} 
}

\maketitle

\begin{abstract}
The rapid advancement of quantum computing (QC) and machine learning (ML) has given rise to the burgeoning field of quantum machine learning (QML), aiming to capitalize on the strengths of quantum computing to propel ML forward. Despite its promise, crafting effective QML models necessitates profound expertise to strike a delicate balance between model intricacy and feasibility on Noisy Intermediate-Scale Quantum (NISQ) devices. While complex models offer robust representation capabilities, their extensive circuit depth may impede seamless execution on extant noisy quantum platforms.
In this paper, we address this quandary of QML model design by employing deep reinforcement learning to explore proficient QML model architectures tailored for designated supervised learning tasks. Specifically, our methodology involves training an RL agent to devise policies that facilitate the discovery of QML models without predetermined ansatz. Furthermore, we integrate an adaptive mechanism to dynamically adjust the learning objectives, fostering continuous improvement in the agent's learning process.
Through extensive numerical simulations, we illustrate the efficacy of our approach within the realm of classification tasks. Our proposed method successfully identifies VQC architectures capable of achieving high classification accuracy while minimizing gate depth. This pioneering approach not only advances the study of AI-driven quantum circuit design but also holds significant promise for enhancing performance in the NISQ era.
\end{abstract}


\section{\label{sec:Indroduction}Introduction}

Quantum computing (QC) holds the potential to revolutionize computational tasks, offering distinct advantages over classical computers \cite{nielsen2010quantum}. The convergence of advancements in quantum hardware and machine learning applications has sparked a growing interest in exploring the synergies between these cutting-edge technologies.
Although existing quantum computers still suffer from noise, a promising solution lies in a hybrid quantum-classical framework. Here, computational tasks are divided into two parts: one executed on a quantum computer and the other on a classical computer \cite{cerezo2021variational, bharti2022noisy}. Central to this paradigm is the Variational Quantum Algorithm (VQA) \cite{cerezo2021variational, bharti2022noisy}, which serves as the cornerstone of hybrid computing. Quantum machine learning (QML) algorithms heavily rely on VQAs, utilizing variational quantum circuits (VQCs) as trainable components akin to classical neural networks.
QML has demonstrated remarkable success across various domains, including classification \cite{mitarai2018quantum, chen2022quantumCNN,chen2021end, l2024quantum,wu2022poster}, time-series modeling \cite{chen2022quantumLSTM}, natural language processing \cite{li2023pqlm,yang2022bert,di2022dawn,stein2023applying}, generative modeling \cite{stein2021qugan, kolle2024quantumDenoisingDiffusionModel,chu2023iqgan}, and reinforcement learning \cite{chen2020variational, jerbi2021parametrized, skolik2022quantum, meyer2022survey,chen2023efficientQRL_QRC,chen2023quantum_LSTM_RL,yun2023quantum,chen2024learning}.
While existing QML models have shown promise, they often require expert knowledge to design effective quantum circuit architectures. For instance, the configuration of encoding and variational subcircuits within VQCs significantly influences model performance and the realization of potential quantum advantages \cite{abbas2021power}. Moreover, the vast search space of VQCs presents a challenge, given the multitude of possible circuit architectures. The necessity for expertise in quantum circuit design poses a barrier to the widespread adoption of QML techniques beyond the quantum computing community, limiting their application in other scientific domains.

\begin{figure}[htbp]
\begin{center}
\includegraphics[width=1\columnwidth]{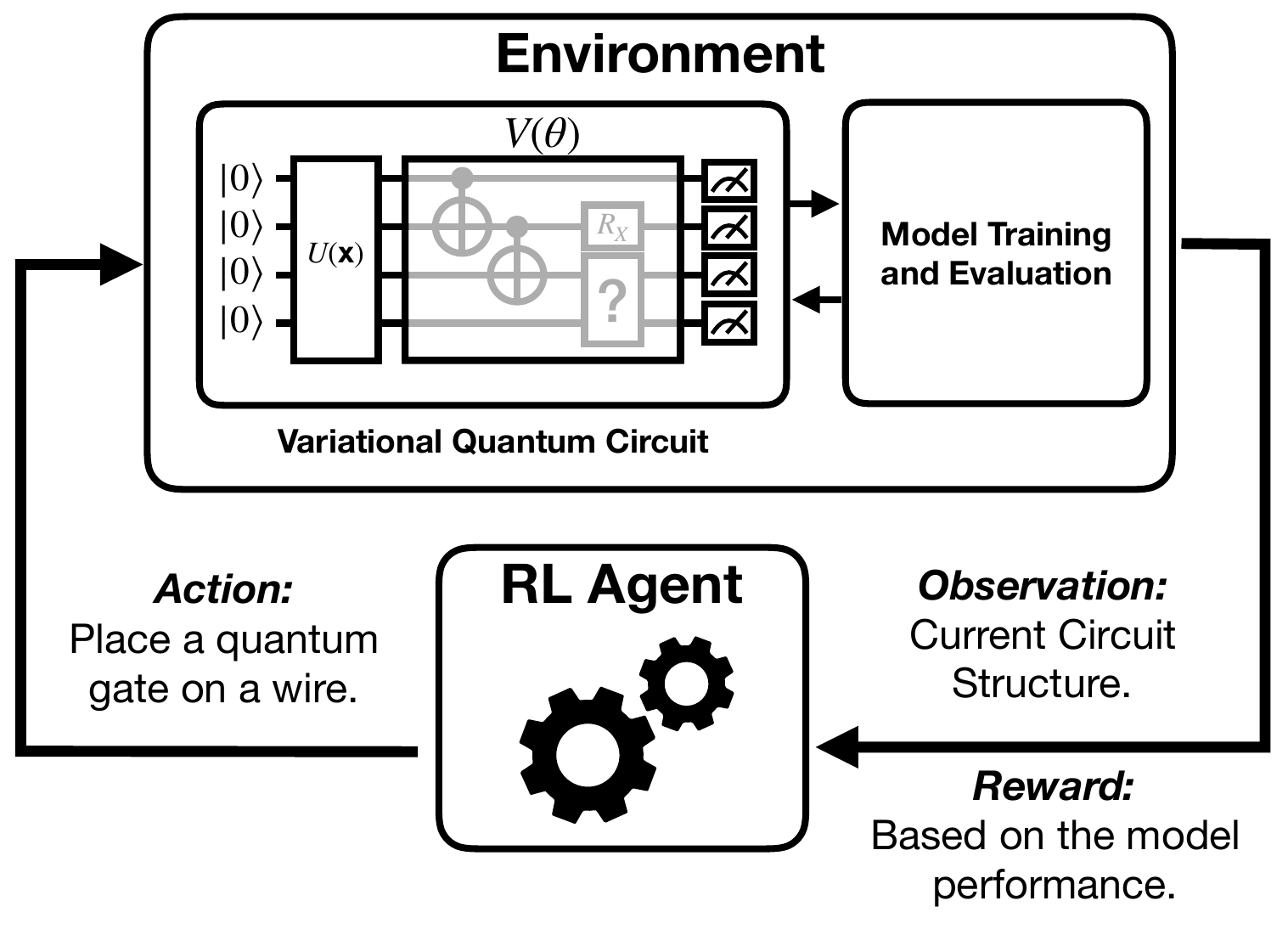}
\caption{{\bfseries Overall scheme for RL-QMLAS.}}
\label{fig:overall_scheme}
\end{center}
\end{figure}

In this paper, we address the challenge of designing QML models by introducing a novel approach called deep reinforcement learning with adaptive search of learning targets (RL-QMLAS). Our method, shown in \figureautorefname{\ref{fig:overall_scheme}}, focuses on classification tasks, wherein the RL agent's objective is to discover an optimal quantum gate sequence. Furthermore, we incorporate an adaptive learning threshold, dynamically adjusting the reward scheme during RL training to enhance the agent's learning process, enabling the model to acquire high-performing policies effectively.
Through numerical simulations, we demonstrate that our proposed methods can effectively generate VQC architectures. These architectures achieve high classification accuracy without requiring prior physical knowledge and maintain a shallower circuit depth compared to manually crafted architectures. 
In addition, the adaptive learning target can enhance the agent learning process and reduce the need for a high pre-defined learning target.
This paper is organized as follows: \sectionautorefname{\ref{sec:relevant_works}} provides a brief survey on current development of QAS. In \sectionautorefname{\ref{sec:VQC}}, we describe the concept of VQC, which is the core of existing QML models and the target the proposed framework is to search for. We formulate the QAS problem in \sectionautorefname{\ref{sec:QAS}} and describe the RL techniques used in this work in \sectionautorefname{\ref{sec:RL}}. We provide the details of simulation in \sectionautorefname{\ref{sec:Methods}} and results in \sectionautorefname{\ref{sec:Results}}. Finally conclude in \sectionautorefname{\ref{sec:Conclusion}}.

\section{\label{sec:relevant_works} Relevant Works}
Machine learning techniques have been applied to tackle various quantum computing challenges such as quantum architecture search (QAS).
The target task of a QAS might be generating a desired quantum state \cite{kuo2021quantum,ye2021quantum,kimura2022quantum,sogabe2022model,lu2023qas,kundu2024enhancing,sunkel2023ga4qco,zhu2023quantum,chen2023QRL_QAS,selig2023deepqprep,sun2024quantum}, finding an efficient circuit for solving chemical ground states \cite{ostaszewski2021reinforcement,wang2023automated,he2023gnn,sun2024quantum,deng2023progressive}, solving an optimization task \cite{yao2022monte,duong2022quantum,wang2023automated,wu2023quantumdarts,sun2024quantum,zhang2022differentiable,sun2024differentiable,liu2023reinforcement}, optimizing a given quantum circuit for a particular hardware architecture \cite{fosel2021quantum}, compiling a circuit \cite{he2022quantum,he2022search,chen2022efficient} or performing a machine learning task \cite{ding2022evolutionary,duong2022quantum,wu2023quantumdarts,zhang2023evolutionary,ding2023multi,subasi2023toward,sun2023differentiable,zhang2021neural,du2022quantum}.
Various approaches are employed to find the optimal circuit for specified tasks. For example, the works \cite{kuo2021quantum,ye2021quantum,fosel2021quantum,ostaszewski2021reinforcement,yao2022monte,kimura2022quantum,sogabe2022model,kundu2024enhancing,zhu2023quantum,chen2023QRL_QAS,chen2022efficient} consider the reinforcement learning based methods while the works \cite{ding2022evolutionary,zhang2023evolutionary,ding2023multi,sunkel2023ga4qco} works use different variants of evolutionary algorithms to search for the circuit. Differentiable QAS methods are also developed to leverage the highly successful gradient-based methods \cite{wu2023quantumdarts,zhang2022differentiable,sun2024differentiable,sun2023differentiable}. Different ways of encoding the quantum circuit architecture are devised. For example, the works \cite{duong2022quantum,he2023gnn} propose graph-based method while the work \cite{fosel2021quantum} consider the convolutional neural network based method to encode the quantum circuit architecture.
Regarding the circuit performance metric, it can be a direct evaluation of the circuit performance on the particular task \cite{ostaszewski2021reinforcement,ding2022evolutionary,wang2023automated} or the closeness of the generated circuit to the actual circuit \cite{kuo2021quantum,ye2021quantum,duong2022quantum}. To reduce the computational resource required in direct evaluation, certain predictor-based methods are proposed to use neural network to predict the quantum model performance without direct circuit evaluation \cite{deng2023progressive,zhang2021neural}.
The proposed method in this paper is to further generalize the concepts used in \cite{kuo2021quantum,ye2021quantum} to more than finding a quantum circuit to synthesize a particular quantum state, but can actually perform a QML task. This paper further generalize the methods proposed in the work \cite{ostaszewski2021reinforcement} to QML tasks. Our work is also different from previous works on quantum circuit optimization \cite{fosel2021quantum}, since in our work, circuit ansatz are not provided. 
Though optimizing an existing circuit ansatz can decrease training time, it is not without cost; the ansatz necessitates input from quantum experts. Our approach investigates the prospect of whether, in the absence of a predefined ansatz, an agent can autonomously discover high-performing circuits.
\section{\label{sec:VQC}Variational Quantum Circuits}
\begin{figure}[htbp]
\begin{center}
\includegraphics[width=1\columnwidth]{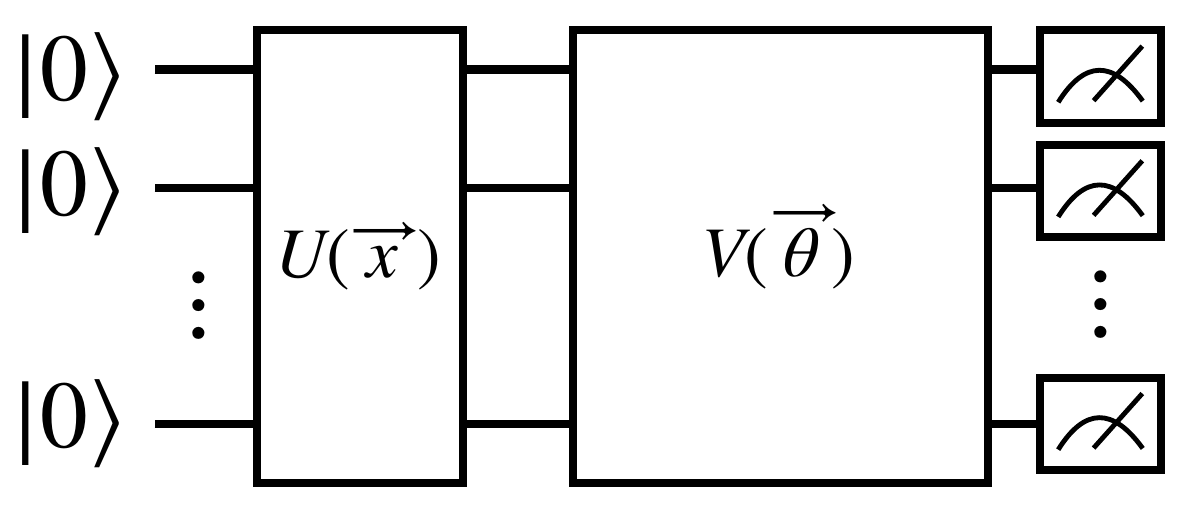}
\caption{{\bfseries Generic variational quantum circuit (VQC) structure.}}
\label{fig:generic_vqc}
\end{center}
\end{figure}

Variational quantum circuits (VQC), also known as parameterized quantum circuits (PQC) is a special kind of quantum circuit with trainable parameters which can be trained via gradient-based \cite{schuld2019evaluating,bergholm2018pennylane,mitarai2018quantum} or gradient-free \cite{chen2022variational} methods. This kind of circuits play a crucial role in the hybrid quantum-classical computing paradigm in which certain computing tasks are implemented on quantum computer while tasks not suitable for existing quantum computers are carried out by classical computers. 
Consider an $n$-qubit system. The fundamental components of a VQC (illustrated in \figureautorefname{\ref{fig:generic_vqc}}) include the \emph{encoding} circuit $U(\vec{x})$, responsible for transforming the classical input vector $\vec{x}$ into a quantum state $U(\vec{x})\ket{0}^{\otimes n}$, the \emph{variational} circuit $V(\vec{\theta})$, serving as the actual learning component with trainable parameters $\vec{\theta}$, and the final \emph{measurement} operation, used to extract information from the circuit.
The VQC used in this work can be expressed as $\overrightarrow{f(\vec{x} ; \vec{\theta})}=\left(\left\langle\hat{Z}_1\right\rangle, \cdots,\left\langle\hat{Z}_n\right\rangle\right)$ , where $\left\langle\hat{Z}_{k}\right\rangle =\left\langle 0\left|U^{\dagger}(\vec{x})V^{\dagger}(\vec{\theta}) \hat{Z_{k}} V(\vec{\theta})U(\vec{x})\right| 0\right\rangle$. The $Z$-expectation values can be derived via multiple sampling (shots) on real quantum devices or direct computation when using a simulation software.
VQCs have been shown to provide certain advantages over classical neural networks \cite{abbas2021power,caro2022generalization} and have demonstrated successful applications in various ML tasks \cite{mitarai2018quantum,chen2022quantumLSTM,chen2022quantumCNN,chen2020variational,jerbi2021parametrized,skolik2022quantum,l2024quantum,kolle2024quantumDenoisingDiffusionModel}. 
The variational circuit $V(\vec{\theta})$ requires special attention since the design of this circuit component will affect the QML model significantly. In general, several control gates and rotation gates are required in this circuit component, and there are several ansatzes which have been shown to be successful. However, these designs are not tailored for specific QML tasks, therefore may not be the optimal in terms of circuit depth. 
\section{\label{sec:QAS}Quantum Architecture Search}
In this paper, we want to solve the following problem: Suppose we are given an initial quantum state $\ket{0} ^{\otimes n}$, an supervised learning dataset $\{(x_{i}, y_{i})\}$ , and encoding circuit $U$, the maximum gate number $L$, the allowed gate set $\mathbb{G}$, an performance metric $\mathcal{M}$ (e.g. classification accuracy or the loss function $\mathcal{L}$), the goal is to find the quantum gate sequence $S$ such that the performance metric is maximized (or minimized). 
\begin{definition}[QAS for Quantum Supervised Learning]
    Given an $n$-qubit system with ground state initialization $\ket{0}^{\otimes n}$ and a predefined encoding circuit $U$, the QAS for quantum supervised learning is to find the gate sequence with length $ < L$ composed from the allowed gate sets $\mathbb{G}$ to build the trainable circuit $V(\vec{\theta})$ such that, after the predefined training process, the quantum function $\overrightarrow{f(\vec{x} ; \vec{\theta})}=\left(\left\langle\hat{Z}_1\right\rangle, \cdots,\left\langle\hat{Z}_m\right\rangle\right)$, where $\left\langle\hat{Z}_{k}\right\rangle =\left\langle 0\left|U^{\dagger}(\vec{x})V^{\dagger}(\vec{\theta}) \hat{Z_{k}} V(\vec{\theta})U(\vec{x})\right| 0\right\rangle$ represents the $Z$ expectation value on $k$-th qubits and $m < n$ equals to the number of outputs, can minimize or maximize the given performance metric $\mathcal{M}(y_{i}, \hat{y}_{i})$. Here the $y_{i}$ and $\hat{y}_{i}$ represent the ground truth and labels predicted by the quantum model, respectively. The predicted label $\hat{y}_{i}$ is derived from the $\overrightarrow{f(\vec{x} ; \vec{\theta})}$ and can be represented as $\hat{y}_{i} = g(\overrightarrow{f(\vec{x} ; \vec{\theta})})$, where $g$ is a post-processing function for $f$.
    Specifically, for our binary classification task we have $\mathcal{M}(y_{i}, \hat{y}_{i}) = -\left( y_i \log(\hat{y}_i) + (1 - y_i) \log(1 - \hat{y}_i) \right)$ and $g = (1 + \overrightarrow{f(\vec{x} ; \vec{\theta})})/2$
\end{definition}
The allowed action (gate) set we consider for this particular problem is
\begin{equation}
\mathbb{G}=\bigcup \limits_{i=1}^{n}  \left\{R_{X_i}, R_{Y_i}, R_{Z_i}, CNOT_{i,(i+1)(mod2)}\right\},
\label{eq:action_gates}
\end{equation}
where $R_{X_i}$, $R_{Y_i}$ and $R_{Z_i}$ represent rotations along $X$, $Y$ and $Z$ axis, respectively.

\section{\label{sec:RL}Reinforcement Learning}

\emph{Reinforcement learning} (RL) stands as a machine learning approach where an \emph{agent} learns decision-making by interacting with environments \cite{sutton2018reinforcement}. In this setup, the \emph{agent} engages with an \emph{environment} $\mathcal{E}$ over discrete time steps. At each time step $t$, the agent receives a current \emph{state} or \emph{observation} $s_t$ from the environment $\mathcal{E}$ and proceeds to select an \emph{action} $a_t$ from a set of available actions $\mathcal{A}$ based on its governing \emph{policy} $\pi$. This policy $\pi$ functions to map the state or observation $s_t$ to the action $a_t$. Typically, the policy may adopt a probabilistic nature, implying that given a state $s$, the action output can be a probability distribution $\pi(a_t|s_t)$ conditioned on $s_t$. Upon executing the action $a_t$, the agent encounters the subsequent state $s_{t+1}$ and receives a single \emph{reward} $r_t$. This iterative process persists until the agent reaches a terminal state or meets predefined termination conditions (e.g., maximum steps allowed). An \emph{episode} refers to the agent's journey from a randomly chosen initial state through to the terminal state or until it satisfies the stopping criteria.

We define the cumulative discounted reward from time step $t$ as $R_t = \sum_{t'=t}^{T} \gamma^{t'-t} r_{t'}$, where $\gamma$ is the discount factor within the range of $(0,1]$. Essentially, $\gamma$ serves as a parameter set by the investigator to influence how future rewards impact decision-making. A higher $\gamma$ assigns greater importance to future rewards, while a lower $\gamma$ leads to more emphasis on immediate rewards, gradually discounting future ones. The agent's objective is to maximize the expected return from each state $s_t$ during the training phase. The \emph{action-value function}, or \emph{Q-value function}, $Q^\pi (s,a) = \mathbb{E}[R_t|s_t = s, a]$ represents the anticipated return for choosing action $a$ in state $s$ according to policy $\pi$. The optimal action-value function $Q^*(s,a) = \max_{\pi} Q^\pi(s,a)$ indicates the highest achievable action-value across all conceivable policies. Furthermore, the value of state $s$ under policy $\pi$, $V^\pi(s) = \mathbb{E}\left[R_t|s_t = s\right]$, represents the agent's expected return when adhering to policy $\pi$ from state $s$. Various reinforcement learning (RL) algorithms aim to identify the policy that maximizes the value function. Algorithms geared towards maximizing the value function are termed \emph{value-based} RL algorithms. One of the notable example of value-based RL is the $Q$-learning \cite{sutton2018reinforcement}.
\subsection{\textit{Q}-Learning}
$Q$-learning \cite{sutton2018reinforcement} stands out as one of the predominant and fundamental model-free algorithms in RL. In $Q$-learning, the agent acquires knowledge of the optimal action-value function and operates as an \emph{off-policy} algorithm. The learning process starts with the random initialization of the value function $Q^{\pi}(s,a)$ for all states $s\in S$ and actions $a\in \mathcal{A}$, typically stored in a structured form known as the $Q$-table. The estimates for $Q^{\pi}(s,a)$ are then progressively updated according to the Bellman equation: 
\begin{align}
  Q\left(s_{t}, a_{t}\right) & \leftarrow  Q\left(s_{t}, a_{t}\right) \nonumber \\
  &+\alpha\left[r_{t}+\gamma \max _{a} Q\left(s_{t+1}, a\right)-Q\left(s_{t}, a_{t}\right)\right].
\end{align}
\subsection{\label{subsec:DDQ}Double Deep Q-Learning}
The conventional $Q$-learning approach, as previously elucidated, offers the theoretically optimal action-value function. However, it becomes impractical for problems necessitating extensive memory. Particularly, managing problems characterized by high-dimensional state ($s$) or action ($a$) spaces poses significant challenges. Moreover, in environments featuring continuous state values, the efficient storage of $Q(s,a)$ within a table is unclear. To circumvent this memory constraint, neural networks (NNs) are employed to effectively represent $Q^{\pi}(s,a) \forall s \in S, a \in \mathcal{A}$. This technique, termed \emph{deep $Q$-learning}, utilizes NNs, with the network itself referred to as a deep $Q$-network (DQN) \cite{mnih2015human}.

To enhance the stability of the deep DQN, methods such as \emph{experience replay} and the integration of an auxiliary network referred to as the \emph{target network} are employed \cite{mnih2015human}. Experience replay involves the agent storing encountered experiences during episodes in memory, preserving transition tuples, denoted as ${s_{t}, a_{t}, r_{t}, s_{t+1}}$. Upon accumulating a sufficient pool of experiences, the agent randomly selects a batch for computation of loss and subsequent update of DQN model parameters. Furthermore, to mitigate the correlation between target and prediction, a duplicate of the DQN, termed the \emph{target network}, is utilized. The parameters $\theta$ of the DQN are updated iteratively, whereas the parameters $\theta^{-}$ of the target network undergo updates at periodic intervals. The DQN training is done via minimizing the mean square error (MSE) loss function:
\begin{equation}
\resizebox{0.9\columnwidth}{!}{
$L(\theta)=\mathbb{E}\left[\left(r_{t}+\gamma \max _{a^{\prime}} Q\left(s_{t+1}, a^{\prime} ; \theta^{-}\right)-Q\left(s_{t}, a_{t} ; \theta\right)\right)^{2}\right]$}
\end{equation}
Other loss functions such as Huber loss or mean absolute error (MAE) can also be used.
Despite the considerable success achieved by DQN, instances arise where it tends to overestimate the action-value function \cite{doubledqn}. As a remedy, an enhanced variant of DQN, known as \emph{Double Deep $Q$-learning} (DoubleDQN), has been devised \cite{doubledqn}. The essence of Double Deep $Q$-learning lies in deconstructing the max operation within the target $y^{DQN}_{t} = r_{t}+\gamma \max_{a^{\prime}} Q\left(s_{t+1}, a^{\prime} ; \theta^{-}\right)$ into two distinct operations: \emph{action selection} and \emph{action evaluation}. Initially, action selection relies on the policy network, $\operatorname{argmax}_{a} Q\left(s_{t+1}, a ; \theta\right)$, followed by the utilization of the target network to evaluate the action, $Q\left(s_{t+1}, \operatorname{argmax}_{a} Q\left(s_{t+1}, a ; \theta\right), \theta^{-}\right)$. Consequently, the DoubleDQN target is reformulated as $y^{DDQN}_{t} = r_{t}+\gamma Q\left(s_{t+1}, \operatorname{argmax}_{a} Q\left(s_{t+1}, a ; \theta\right), \theta^{-}\right)$.

The loss function $L(\theta)$ is therefore:

\begin{align}
    L(\theta)&=\mathbb{E}\Bigl[\left(r_{t} +  \gamma Q\left(s_{t+1}, \operatorname{argmax}_{a} Q\left(s_{t+1}, a ; \theta\right), \theta^{-}\right)  \right.  \nonumber \\
    & \qquad \qquad   \left. -Q\left(s_{t}, a_{t} ; \theta\right)\right)^{2}\Bigr] 
\end{align}

Then, $\theta$ is updated using the gradient descent method and every few iterations we update the target network $\theta^{-}\leftarrow\theta$.
\subsection{\label{subsec:N_Step_DDQN}N-Step DDQN}
The $N$-step DDQN extends the standard DDQN by considering a sequence (trajectory) of $N$ steps when updating the $Q$-values according to the following loss function,
\begin{align}
    &L(\theta)=\mathbb{E}\Biggl[\Biggl(\sum_{k=0}^{N-1} \gamma^k r_{t+k+1} +\gamma^N \nonumber \\ 
    &Q\left(s_{t+N}, \operatorname{argmax}_{a} Q\left(s_{t+N}, a ; \theta\right), \theta^{-}\right) -Q\left(s_{t}, a_{t} ; \theta\right)\Biggr)^{2}\Biggr]
\end{align}
where $\gamma$ represents the discount factor, $\theta$ represents the policy net parameters, $\theta^{-}$ represents the target net parameters and $r_{t+k+1}$ is the reward received at timestep $t+k+1$. Note that here we use MSE loss as an example, however other kind of loss function can be used to fine-tune the model performance. In this work, we use the \texttt{Smooth\_L1} loss.
By considering multiple steps, the $N$-step DDQN provides a more informative signal for updates and allows the agent to consider the long-term consequences of its actions, potentially leading to faster convergence and improved performance compared to the standard DQN and DDQN.

\section{\label{sec:Methods}Methods}

In this work, we use the \textsc{TensorCircuit}~\cite{zhang2022tensorcircuit} for constructing the variational quantum circuits and \textsc{PyTorch}~\cite{paszke2019pytorch} for building the deep reinforcement learning model.

\subsection{Experimental setup}
In our experimental setup, we employ two types of datasets from \textsc{scikit-learn}~\cite{scikit-learn} to generate data for the binary classification task. The primary dataset is generated using \texttt{sklearn.datasets.make\_classification}. 
This function creates $n$-dimensional datasets where data points form normally distributed clusters (with a standard deviation of 1) around the vertices of an $n_{\rm informative}$-dimensional hypercube. The remaining $n_{\rm redundant} = n - n_{\rm informative}$ features are random linear combinations of these informative features, adding complexity to the classification task. 
In addition to the \texttt{make\_classification} dataset, we also utilize the \texttt{sklearn.datasets.make\_moons} dataset. This dataset generates two interleaving half-moon shaped clusters, which are particularly useful for evaluating the model's ability to capture non-linear decision boundaries in binary classification tasks.

For each episode within our RL-QMLAS framework, we initialize the environment with an empty quantum circuit which is in the ground state $\ket{0}^{ \otimes n}$. At every step, the agent selects a quantum gate and its location based on the output from the policy network. The set of permissible quantum gates includes rotation gates ($R_{X}$, $R_{Y}$ and $R_{Z}$) and the CNOT gate. 
To maintain a manageable action space, the agent is tasked with choosing the type of the rotation gate (either $X$, $Y$, or $Z$), while the specific rotation angle is optimized through a classical optimizer during the training of the quantum classifier. This approach balances the complexity of quantum gate selection with practical training considerations.
Input data is encoded into the quantum circuit using an \texttt{arctan} embedding strategy, specifically, for each feature vector $f \in \mathbb{R}^n$, we compute angles $\theta_i = \arctan(f_i)$ and $\phi_i = \arctan(f_i^2)$ for $i \in \{1, 2, \ldots, n\}$. These angles are then used to apply rotation gates as follows:
\begin{equation}\label{eq:artan_embedding}
\forall i \in \{1, 2, \ldots, n\}, \quad \text{apply } R_{Y}(\theta_i) \text{ and } R_{Z}(\phi_i) \text{ on qubit } i.
\end{equation}
Upon the addition of a new gate to the circuit, the quantum classifier is trained for a fixed number of epochs, or until it reaches the desired accuracy level. This iterative process of gate selection and classifier training continues, evolving the quantum circuit step-by-step until the episode concludes, either by achieving the desired accuracy or by reaching the maximum limit of quantum gates. See \sectionautorefname{\ref{subsubsec:reward-function}} for details about the reward scheme.

Significantly, this method of quantum architecture search through reinforcement learning negates the need for prior physical knowledge, enabling the algorithm to autonomously discover efficient and effective quantum circuits. It represents a novel approach where the intricacies of quantum computation are navigated and optimized through machine learning, rather than relying on pre-established physical \emph{ansatz}.

\subsection{Hyperparameters of RL }
\subsubsection{N-step Double Deep Q-Network (DDQN)}
In this study, we applied an $N$-step Double Deep Q-Network (DDQN) \cite{doubledqn} to learn efficient quantum circuits for classification tasks.
The discount factor $\gamma$ is set as $\gamma = 0.005^{1/L}$, where $L$ represents the maximum number of quantum gates allowed, promoting an approach that favors achieving tasks with minimal gate usage. To stabilize the learning process, we periodically synchronize the parameters of the target network with those of the policy network every 512 steps. An experience replay buffer with a capacity of 16384  transitions is used to break the correlation of sequential learning updates and enhance learning efficiency. 
The exploration strategy is governed by an $\epsilon$-greedy policy, with $\epsilon$ decaying from 1 to 0.1 over time to balance exploration and exploitation.

The architecture of the deep $Q$-network is a multilayer perceptron (MLP) consisting of a sequence of linear layers, each followed by a LeakyReLU activation function and dropout regularization. The input to the MLP is the state (observation) vector, which is a $4\times L$ matrix representing the configuration of the quantum circuit at each step, where $L$ is the maximum number of layers. The first two elements of the state vector denote the locations of the control and NOT gates, while the third and fourth elements indicate the location of the rotation gate and the rotation axis, respectively. 
During training, the DDQN receives a flattened state vector. After each testing episode, the test accuracy is appended to the state vector, providing an additional input feature for the MLP. The output of the MLP corresponds to the $Q$-values for each action, guiding the agent's decision-making process in selecting the most promising gate configurations to explore.

\subsubsection{Reward Function}\label{subsubsec:reward-function}
The RL agent interacts with a quantum circuit environment, where the agent's actions involve selecting the control and target qubits for CNOT gates and the qubit and axis for rotation gates. The environment calculates the accuracy of the resulting quantum circuit and provides a reward signal based on the change in accuracy and the number of layers used. The reward function is defined as follows:
\begin{align}\label{reward_func}
R(l) &= \left\{
    \begin{array}{l}
        0.2 \cdot \left(\frac{y_l}{y_{\rm target}} \right)  \cdot (L - l), \text{if } y_l \geq y_{\rm target} \text{ and } l < L,\\
        -0.2\cdot \left(\frac{y_{\rm target} - y_l}{y_{\rm target}}\right)  \cdot l, \text{if } y_l < y_{\mathrm{min}} \text{ and } l = L, \\
        \text{clip}\left(\frac{y_l - y_{l-1}}{y_{l-1} + 1 \times 10^{-6}} - 0.01 \cdot l, -1.5, 1.5\right), \text{otherwise.}
    \end{array}
\right.
\end{align}
Here $y_{\rm{target}}$ is the target accuracy. The reward function encourages the agent to achieve or surpass the target accuracy with minimal gate usage. The scaling factor 0.2 moderates the reward magnitude to ensure stability. Moreover, the function imposes a penalty if $y_{\rm{target}}$ is not reached when the maximum number of gates is used. \equationautorefname{\ref{reward_func}} also dynamically rewards small improvements in accuracy (the 3rd line), but this reward decreases as more gates are added, steering the agent towards more efficient solutions. To maintain numerical stability and prevent extreme values from skewing the agent's learning, the dynamic reward is constrained within the range $[-1.5, 1.5]$. This careful design of the reward function ensures an optimal trade-off between accuracy and efficiency.

\subsubsection{Adaptive Search}
One potential drawback of our previous approach is that the desired classification accuracy $y_{\rm target}$ is determined \emph{a priori}. 
If the $y_{\rm target}$ is set too high, the agent will likely to fail in most of the cases and we need to manually increase $L$, which leads to slow convergence. 
On the other hand, a $y_{\rm target}$ that's set too low allows the agent to quickly find simplistic solutions, hindering the development of more efficient and sophisticated quantum circuits.

To address the need for pre-selecting an appropriate classification accuracy target ($y_{\rm target}$), we introduce an adaptive search strategy. This approach dynamically adjusts both $y_{\rm target}$ and the exploration rate $\epsilon$ based on the agent's ongoing performance.
During the training phase, $y_{\rm target}$ increases incrementally by 0.01 whenever the agent consistently meets or exceeds this threshold across a specified number of episodes, such as 10 successes in 12 consecutive episodes. 
In the testing phase, a similar mechanism is in place. If the agent repeatedly achieves higher accuracies over 5 consecutive tests, the $y_{\rm target}$ is further increased by 0.01, challenging the agent to refine its performance. Concurrently, $\epsilon$ is decreased to 95\% of its value, shifting the agent's focus from exploration to exploitation. 
As a result, the agent increasingly relies on learned behaviors and experiences rather than random exploration, enhancing its proficiency.

\section{\label{sec:Results}Results}
\subsection{Fixed target results}
\begin{figure}[ht]
\centering
\includegraphics[width=1\columnwidth]{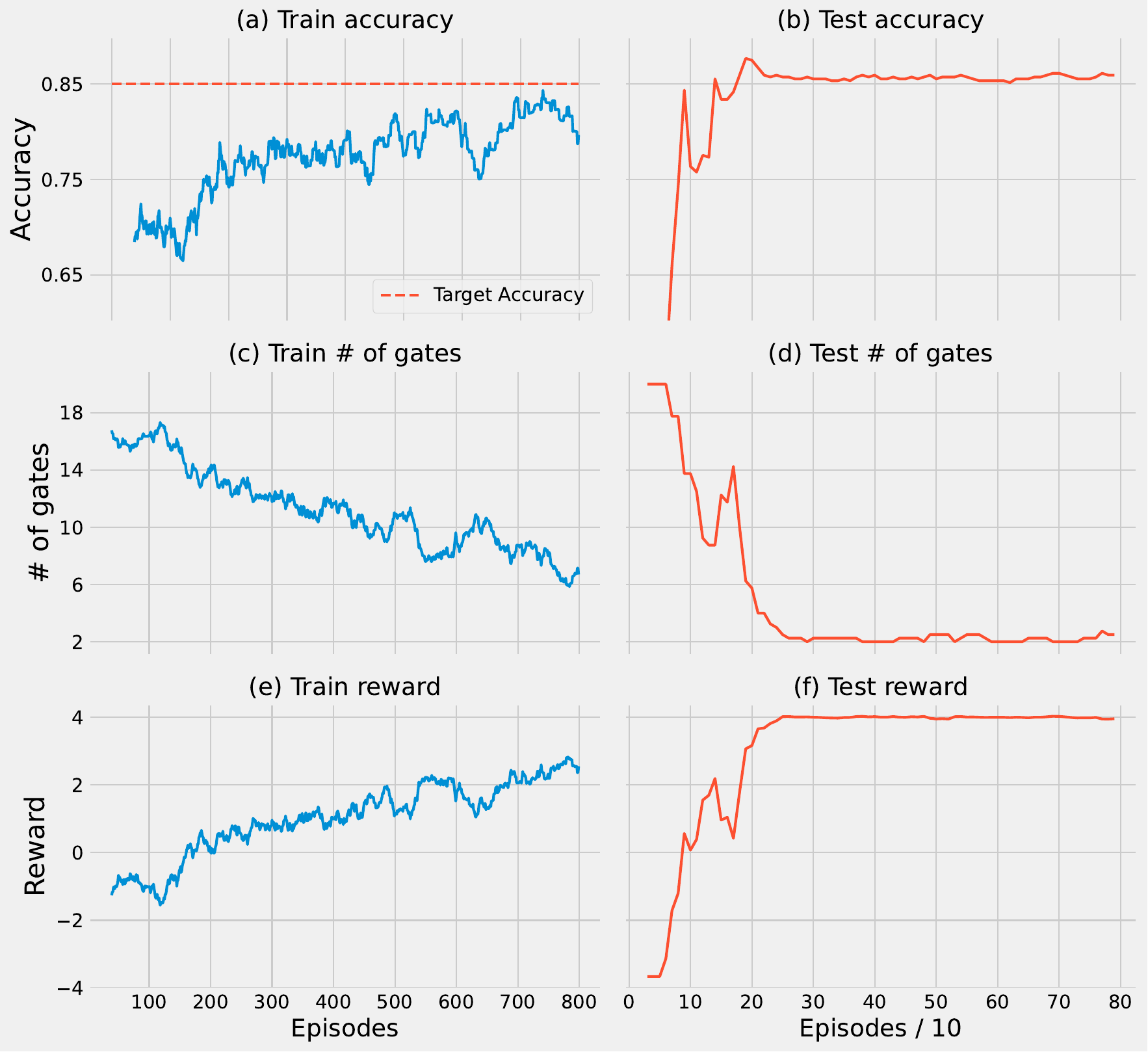}
\caption{Reinforcement learning performance metrics using the \texttt{make\_classification} dataset with a fixed target accuracy of 0.85 and a maximum of 20 quantum gates for a total of 800 episodes. Training accuracy (a) and the number of gates (c) are smoothed with a 40-episode moving average. For testing accuracy (b) and gate count (d), a 4-episode moving average is applied. The maximum training epoch after each action is set to 15. The rewards patterns during training (e) and testing (f) further demonstrate the agent's learning.}
\label{fig:make_classification_fixed_min_acc_85}
\end{figure}

\begin{figure}[ht]
\centering
\includegraphics[width=1\columnwidth]{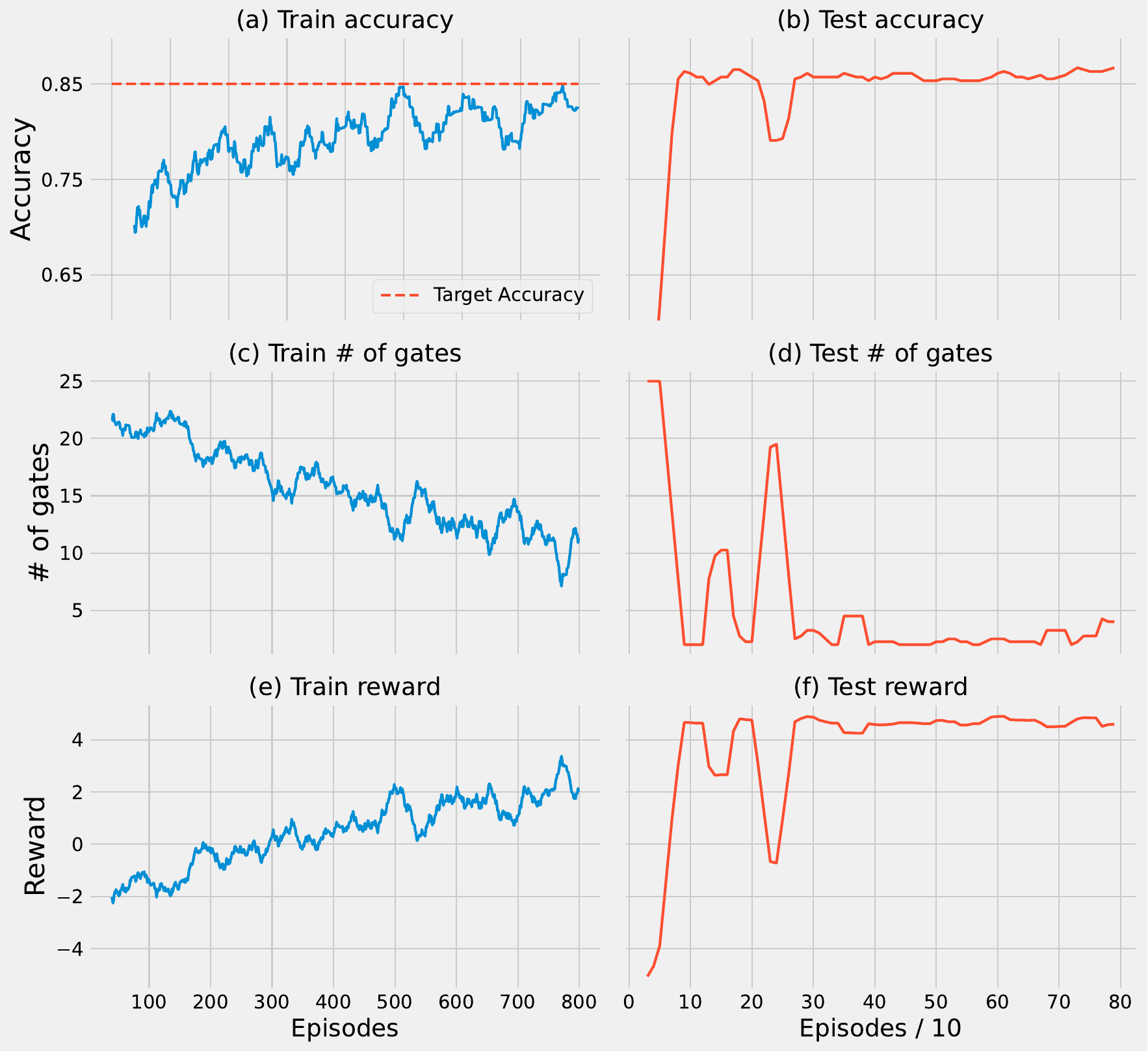}
\caption{
Reinforcement learning performance metrics using the \texttt{make\_moons} dataset with a target accuracy of 0.85 and a maximum of 25 quantum gates for a total of 800 episodes. Training accuracy (a) and the number of gates (c) are smoothed with a 40-episode moving average. For testing accuracy (b) and gate count (d), a 4-episode moving average is applied. The maximum training epoch after each action is set to 25.
}
\label{fig:make_moons_fixed_min_acc_85}
\end{figure}

Our RL agent was first evaluated using the \texttt{make\_classification} dataset. The outcomes are illustrated in \figureautorefname{\ref{fig:make_classification_fixed_min_acc_85}}, where we observed a consistent improvement in classification accuracy (\figureautorefname{\ref{fig:make_classification_fixed_min_acc_85}}a) and a decrease in the number of quantum gates required (\figureautorefname{\ref{fig:make_classification_fixed_min_acc_85}}c). 
In the testing phase, the exploration rate \(\epsilon\) was set to zero, ensuring that the agent's gate selection was entirely based on the learned policy. Notably, a stable and high testing accuracy (\figureautorefname{\ref{fig:make_classification_fixed_min_acc_85}}b) coupled with a minimal number of gates (\figureautorefname{\ref{fig:make_classification_fixed_min_acc_85}}d) was achieved after several hundred training episodes. This indicates the agent's capacity to learn and its efficiency in converging to an optimized quantum circuit structure over the course of training. Moreover, the rewards pattern in both training (\figureautorefname{\ref{fig:make_classification_fixed_min_acc_85}}e) and testing (\figureautorefname{\ref{fig:make_classification_fixed_min_acc_85}}f) further confirmed the agent's proficiency in balancing classification accuracy with circuit simplicity.
For the \texttt{make\_moons} dataset, similar trends were observed (\figureautorefname{\ref{fig:make_moons_fixed_min_acc_85}}), signifying the robustness of our reinforcement learning approach. The agent not only maintained high classification accuracy but also continued to design quantum circuits with an optimized number of gates. This consistency across different datasets indicates the potential of our method to be applied broadly in quantum machine learning tasks without prior physical knowledge.

\subsection{Adaptive Search Results}
The adaptive search strategy has been implemented to dynamically adjust the target accuracy $y_{\rm target}$ and exploration rate $\epsilon$ during the training of our RL agent. This approach is designed to progressively challenge the agent, enhancing its ability to construct efficient quantum circuits. The results of this strategy are discussed below for two datasets.
\begin{figure}[ht]
\centering
\includegraphics[width=1\columnwidth]{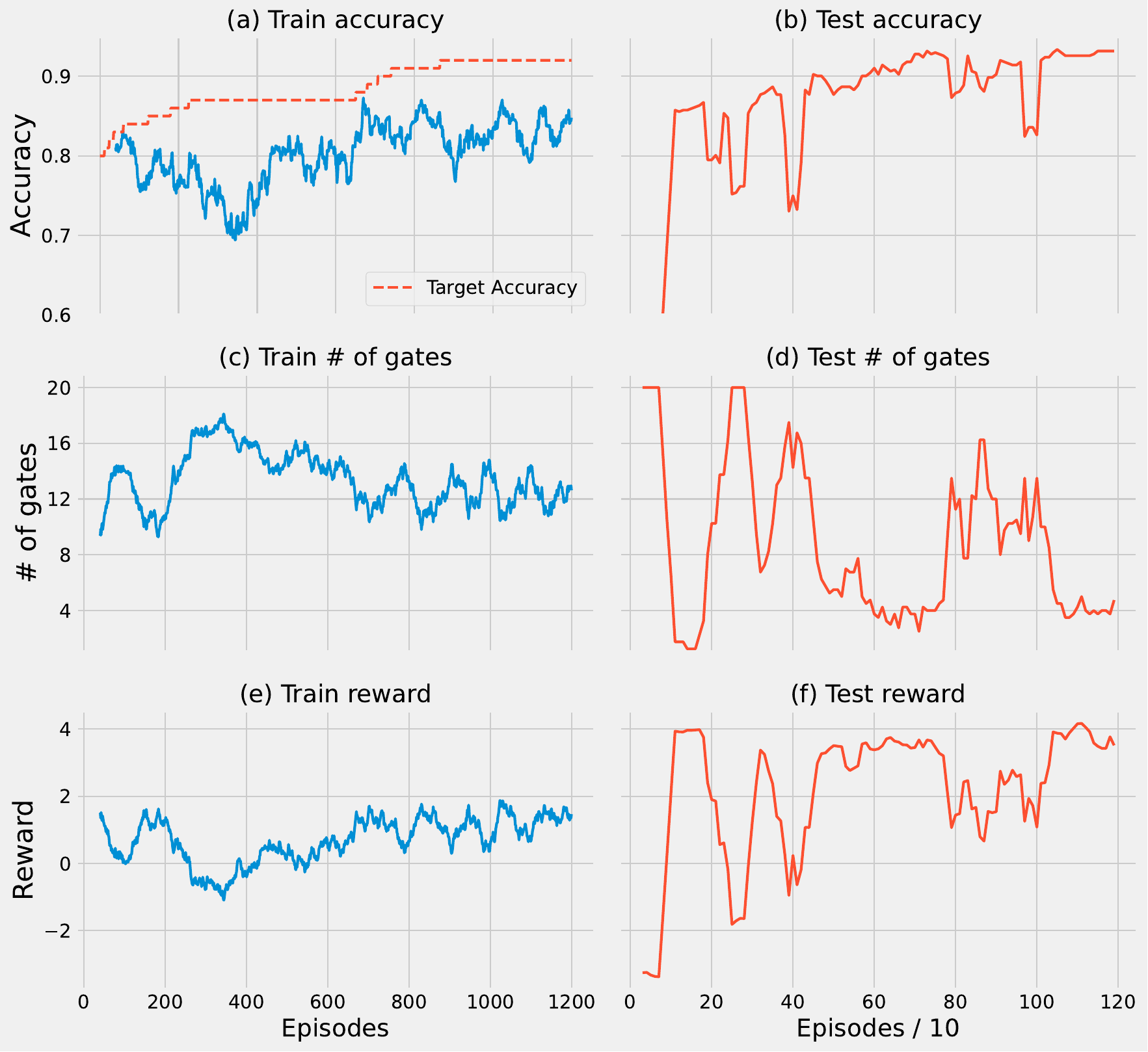}
\caption{
Performance of the reinforcement learning agent on the \texttt{make\_classification} dataset over 1200 episodes using a adaptive search strategy, starting from an initial target accuracy of 0.8. Panel (a) includes the dynamic target accuracy, which is adjusted by the adaptive search strategy, alongside the training accuracy. All other experimental conditions and metric smoothing methods align with those described for the fixed target experiments.}
\label{fig:make_classification_min_acc_80_moving}
\end{figure}

\begin{figure}[ht]
\centering
\includegraphics[width=1\columnwidth]{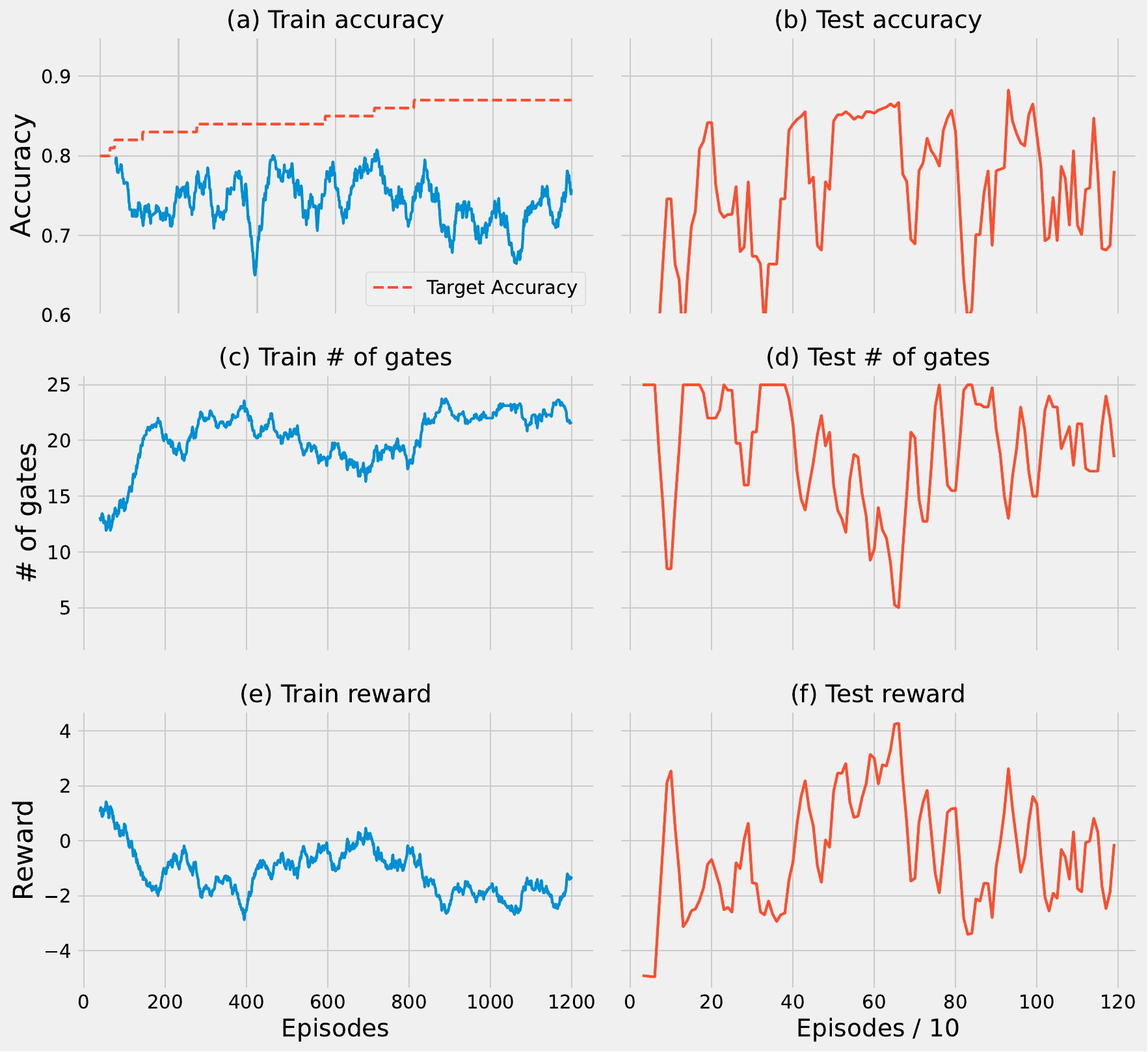}
\caption{Performance of the reinforcement learning agent on the \texttt{make\_moons} dataset over 1200 episodes using a adaptive search strategy, starting from an initial target accuracy of 0.8. Panel (a) includes the dynamic target accuracy, which is adjusted by the adaptive search strategy, alongside the training accuracy. All other experimental conditions and metric smoothing methods align with those described for the fixed target experiments.}
\label{fig:make_moons_min_acc_80_moving}
\end{figure}

\figureautorefname{\ref{fig:make_classification_min_acc_80_moving}} presents the results for the \texttt{make\_classification} dataset. In \figureautorefname{\ref{fig:make_classification_min_acc_80_moving}}a, the agent's training accuracy oscillates around the dynamic target accuracy, $y_{\rm target}$, demonstrating continuous adjustment and learning in response to its evolution. 
To improve the stability and performance, we increased the training episodes from 800 (fixed $y_{\rm target}$ scenario, \figureautorefname{\ref{fig:make_classification_fixed_min_acc_85}}) to 1200. \figureautorefname{\ref{fig:make_classification_min_acc_80_moving}}c shows an initial increase in the number of gates, followed by stabilization, indicating the agent's attempt to balance the quantum circuit's complexity and efficiency. 
During the testing phase, as depicted in \figureautorefname{\ref{fig:make_classification_min_acc_80_moving}}b and d, we observe more significant performance fluctuations compared to the fixed $y_{\rm target}$ scenario in \figureautorefname{\ref{fig:make_classification_fixed_min_acc_85}}. Each increase in $y_{\rm target}$ results in a temporary performance dip, followed by stabilization and alignment with the new $y_{\rm target}$. 
After 1200 episodes, the agent efficiently utilizes 4 quantum gates to achieve a classification accuracy of 0.93.

\begin{figure}[ht]
\centering
\scalebox{1}{
\Qcircuit @C=1em @R=1em {
& \lstick{q_0:} & \gate{RY(\theta_0)} & \gate{RZ(\phi_0)} & \qw & \targ & \qw & \qw \\
& \lstick{q_1:} & \gate{RY(\theta_1)} & \gate{RZ(\phi_1)} & \qw & \qw & \ctrl{1} & \qw \\
& \lstick{q_2:} & \gate{RY(\theta_2)} & \gate{RZ(\phi_2)} & \gate{RY(\phi)} & \ctrl{-2} & \targ & \qw \\
& \lstick{q_3:} & \gate{RY(\theta_3)} & \gate{RZ(\phi_3)} & \qw & \qw & \qw & \qw \\
}
}
\caption{An example of quantum circuit learned by the RL agent for the \texttt{make\_classification} dataset using the adaptive search strategy, corresponding to the experiment shown in \figureautorefname{\ref{fig:make_classification_min_acc_80_moving}} at episode 1200. The state vector for this circuit is $[[4, 0, 2, 2], [2, 0, 4, 0], [1, 2, 4, 0], \ldots]$, where $\ldots$ denotes $[0, 0, 0, 0]$ for the remaining $L-3$ layers, and $L$ is the maximum number of layers allowed. The initial $R_Z$ and $R_Y$ gates correspond to the \texttt{arctan} data embedding (\equationautorefname{\ref{eq:artan_embedding}}).}
\label{fig:quantum_circuits}
\end{figure}
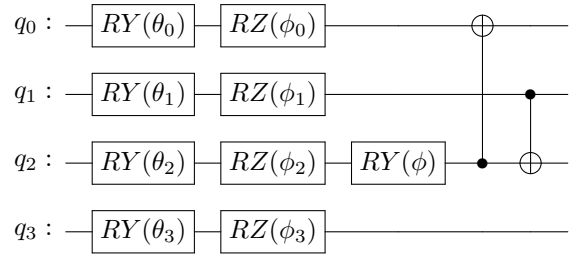

\figureautorefname{\ref{fig:quantum_circuits}} showcases an example of quantum circuit discovered by the RL agent for the \texttt{make\_classification} dataset using the adaptive search strategy, corresponding to the experiment shown in \figureautorefname{\ref{fig:make_classification_min_acc_80_moving}}. This circuit demonstrates the agent's capability to learn and optimize a quantum circuit tailored to the specific task requirements.

For the \texttt{make\_moons} dataset, shown in \figureautorefname{\ref{fig:make_moons_min_acc_80_moving}}, the agent's performance is more variable due to the dataset's inherent complexity. Both training and testing phases indicate greater challenges in consistently meeting the dynamic target accuracy $y_{\rm target}$, especially after it surpasses 0.85 (\figureautorefname{\ref{fig:make_moons_min_acc_80_moving}}b). This suggests a limitation in the agent's capability given the current resources, i.e., the number of qubits and the maximum circuit depth. Nonetheless, the agent shows adaptability and learning capacity, albeit with a less stable optimization process compared to the \texttt{make\_classification} dataset.

\subsection{Comparison with classical machine learning}
A comparative analysis with classical machine learning methods further elucidates the effectiveness of our quantum classifier. We conducted experiments using logistic regression (LR) and support vector machine (SVM) on the same datasets to benchmark performance.

For the \texttt{make\_classification} dataset, both LR and SVM models achieved an accuracy of 90.3\%. In contrast, for the \texttt{make\_moons} dataset, the LR model attained an accuracy of 82.5\%, while the SVM model, leveraging its RBF kernel, reached an accuracy of 100\%.  Notably, the 4-dimensional \texttt{make\_classification} dataset highlights the efficiency of our quantum classifier. With merely 2 quantum gates, it paralleled the performance of the LR and SVM models. The LR model utilized 5 parameters (comprising 4 coefficients and 1 intercept), whereas the SVM model employed a total of 110 support vectors in this instance. In the case of the 2-dimensional \texttt{make\_moons} dataset, the LR model required 3 parameters, and the SVM model used 44 support vectors. Here again, our quantum classifier demonstrated comparable accuracy, albeit with fewer parameters. This outcome is particularly noteworthy for the \texttt{make\_classification} dataset, underscoring the potential of quantum classifiers in achieving high performance with reduced parameterization.

\section{\label{sec:Discussion}Discussion}
While the adaptive search strategy enables the agents to achieve
better accuracy adaptively without relying on the initial guessing, the successful outcome of such strategy still relies partially on the specific target accuracy update scheme. 
For instance, if the target accuracy is increased too quickly, it is possible that the agents will get stuck and the resulting accuracy become worse. Further investigation will be required to establish the optimal schedule of changing the learning target as well as the connection between model performance and the complexity of the dataset.
In more realistic scenarios, quantum devices are subject to noise and decoherence. Consequently, quantum circuit architectures identified through noise-free simulations may exhibit poor performance when directly implemented on real quantum devices. It is therefore compelling to explore the generalization of the proposed RL-QMLAS framework to noisy quantum devices or more challenging conditions, such as fluctuating or drifting noise patterns.
Another direction of investigation involves the scaling behavior of the proposed framework as the number of qubits increases. It is noted that as the number of qubits $n$ increases, the potential combinations of quantum gates scale up rapidly. For example, the number of allowed gates described in \equationautorefname{\ref{eq:action_gates}} scales at $3n + n(n-1) = \Omega(n^{2})$. This rapid scaling presents significant challenges in designing the deep neural networks for the RL agents, as the number of output neurons would increase quickly. Further studies are required to design more efficient RL agents to generate plausible actions for constructing larger-scale quantum circuits.
\section{\label{sec:Conclusion}Conclusion}
In this paper, we present a framework that leverages deep reinforcement learning to construct quantum machine learning models tailored for classification tasks. Through extensive numerical simulations across various scenarios, our approach demonstrates the capability to develop high-performing QML models without the need for manually designing the learnable circuit based on prior physical knowledge. Furthermore, our models achieve commendable performance while utilizing a moderate number of quantum gates, making them suitable for implementation on existing noisy quantum devices. These findings offer a pioneering avenue for exploring the potential of automated QML in diverse application domains.

\begin{figure}[htbp]
\centering
\includegraphics[width=1\columnwidth]{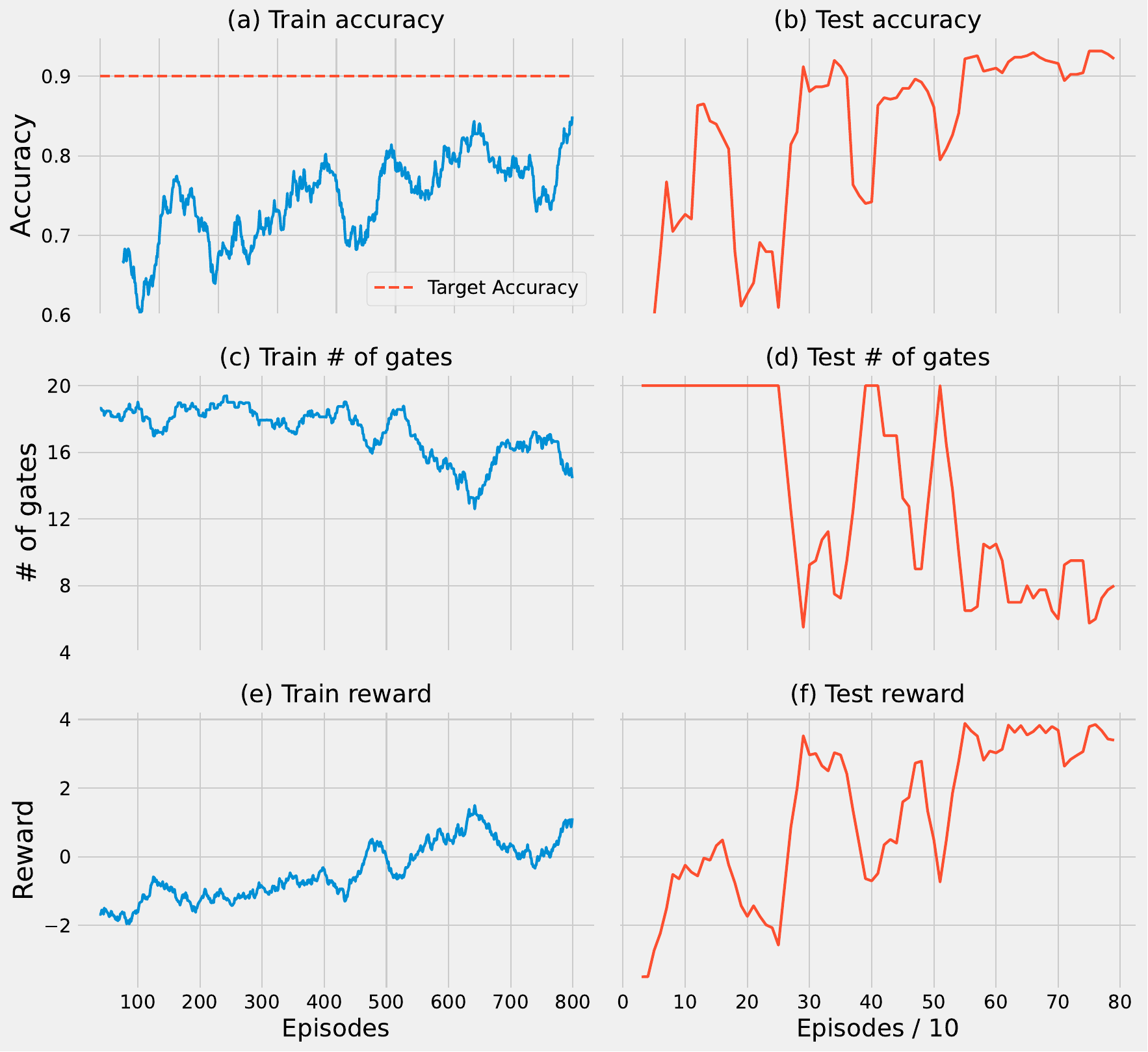}
\caption{Experimental results under the same conditions as in \figureautorefname{\ref{fig:make_classification_fixed_min_acc_85}}, with the only difference being the target accuracy set at 0.90.}\vskip -0.2in
\label{fig:make_classification_fixed_min_acc_90}
\end{figure}

\begin{figure}[htbp]
\centering
\includegraphics[width=1\columnwidth]{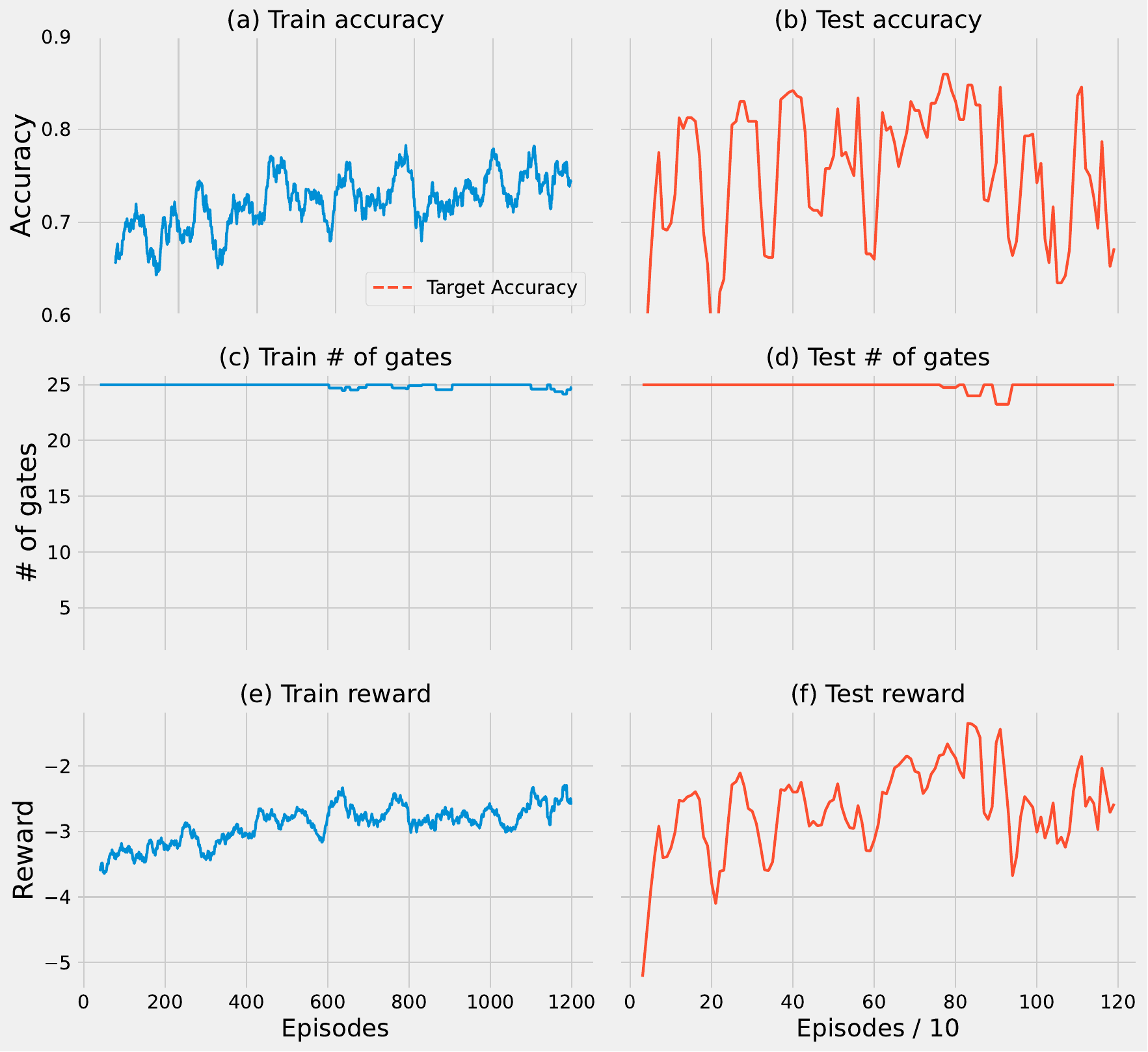}
\caption{Experimental results under the same conditions as in \figureautorefname{\ref{fig:make_moons_fixed_min_acc_85}}, with the only difference being the target accuracy set at 0.90.}\vskip -0.2in
\label{fig:make_moons_fixed_min_acc_90}
\end{figure}

\begin{figure}[htbp]
\centering
\includegraphics[width=1\columnwidth]{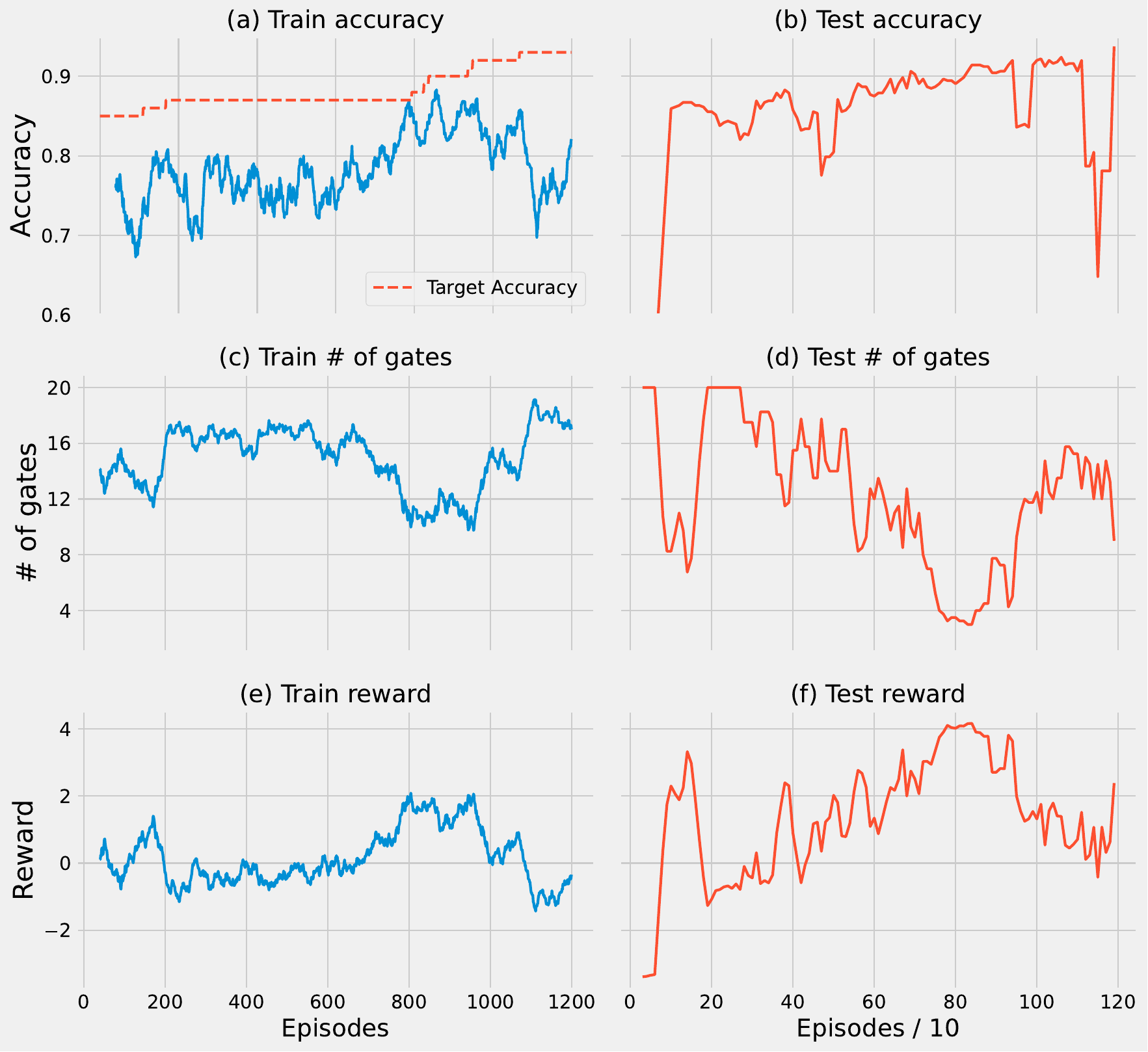}
\caption{Experimental results under the same conditions as in \figureautorefname{\ref{fig:make_classification_min_acc_80_moving}}, with the only difference being the initial target accuracy set at 0.85.}\vskip -0.2in
\label{fig:make_classifications_moving_0.85}
\end{figure}

\begin{figure}[htbp]
\centering
\includegraphics[width=1\columnwidth]{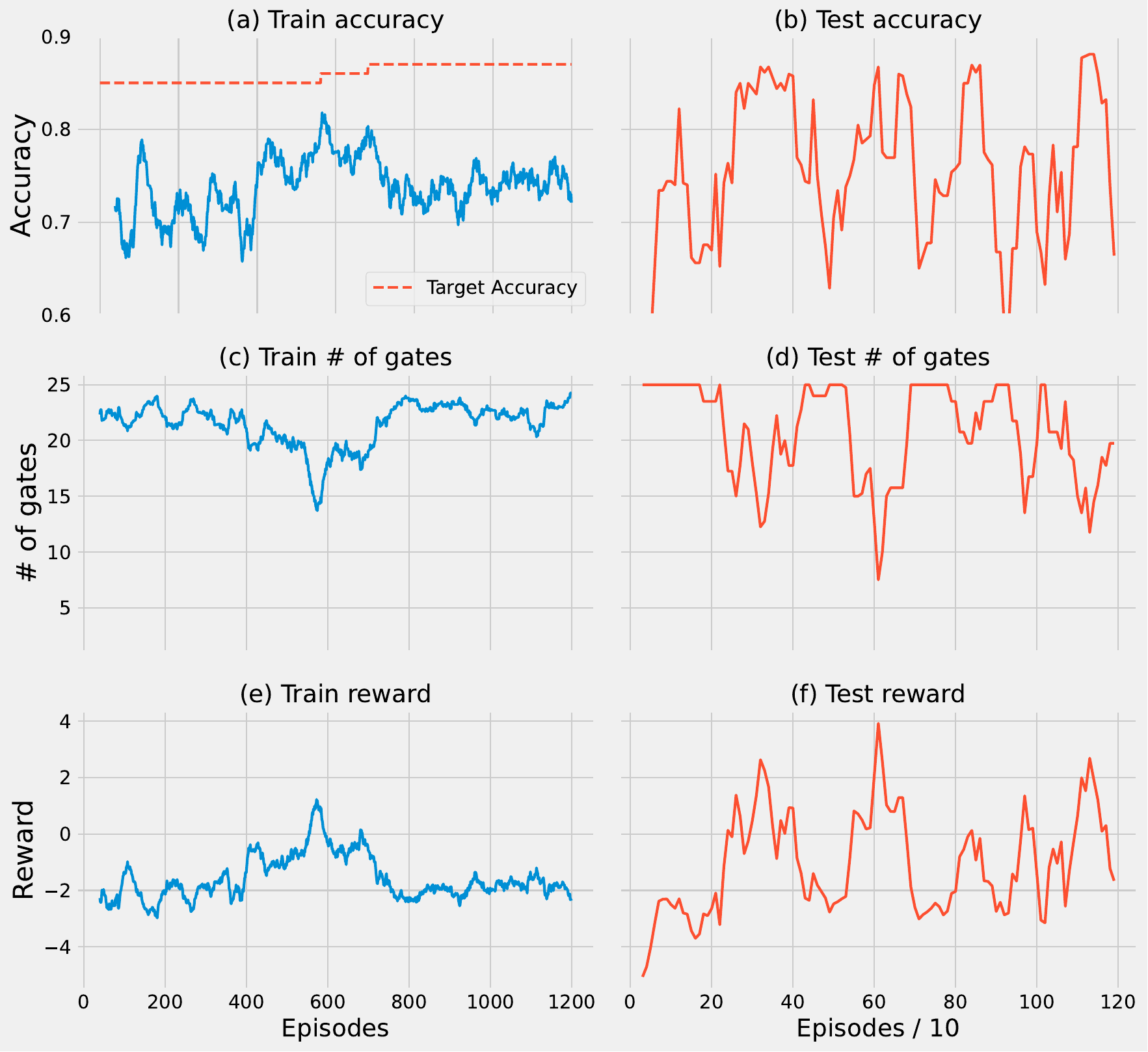}
\caption{Experimental results under the same conditions as in \figureautorefname{\ref{fig:make_moons_min_acc_80_moving}}, with the only difference being the initial target accuracy set at 0.85.}\vskip -0.2in
\label{fig:make_moons_moving_0.85}
\end{figure}

\appendix
In this appendix, we present additional experimental results using the \texttt{make\_classification} and \texttt{make\_moons} datasets under varied conditions, as extensions to the main experiments described in the manuscript. These supplementary results Figs.~\ref{fig:make_classification_fixed_min_acc_90},~\ref{fig:make_moons_fixed_min_acc_90},~\ref{fig:make_classifications_moving_0.85} and \ref{fig:make_moons_moving_0.85} complement the primary findings by illustrating how variations in target accuracies and the implementation of adaptive search strategy influence the agent's learning trajectory and quantum circuit optimization.

\bibliographystyle{ieeetr}
\bibliography{apssamp,bib/qml_examples,bib/qas,bib/qc,bib/classical_rl}

\providecommand{\noopsort}[1]{}\providecommand{\singleletter}[1]{#1}%
\begin{thebibliography}{10}

\bibitem{nielsen2010quantum}
M.~A. Nielsen and I.~L. Chuang, {\em Quantum computation and quantum information}.
\newblock Cambridge university press, 2010.

\bibitem{cerezo2021variational}
M.~Cerezo, A.~Arrasmith, R.~Babbush, S.~C. Benjamin, S.~Endo, K.~Fujii, J.~R. McClean, K.~Mitarai, X.~Yuan, L.~Cincio, {\em et~al.}, ``Variational quantum algorithms,'' {\em Nature Reviews Physics}, vol.~3, no.~9, pp.~625--644, 2021.

\bibitem{bharti2022noisy}
K.~Bharti, A.~Cervera-Lierta, T.~H. Kyaw, T.~Haug, S.~Alperin-Lea, A.~Anand, M.~Degroote, H.~Heimonen, J.~S. Kottmann, T.~Menke, {\em et~al.}, ``Noisy intermediate-scale quantum algorithms,'' {\em Reviews of Modern Physics}, vol.~94, no.~1, p.~015004, 2022.

\bibitem{mitarai2018quantum}
K.~Mitarai, M.~Negoro, M.~Kitagawa, and K.~Fujii, ``Quantum circuit learning,'' {\em Physical Review A}, vol.~98, no.~3, p.~032309, 2018.

\bibitem{chen2022quantumCNN}
S.~Y.-C. Chen, T.-C. Wei, C.~Zhang, H.~Yu, and S.~Yoo, ``Quantum convolutional neural networks for high energy physics data analysis,'' {\em Physical Review Research}, vol.~4, no.~1, p.~013231, 2022.

\bibitem{chen2021end}
S.~Y.-C. Chen, C.-M. Huang, C.-W. Hsing, and Y.-J. Kao, ``An end-to-end trainable hybrid classical-quantum classifier,'' {\em Machine Learning: Science and Technology}, vol.~2, no.~4, p.~045021, 2021.

\bibitem{l2024quantum}
R.~L'Abbate, A.~D'Onofrio, S.~Stein, S.~Y.-C. Chen, A.~Li, P.-Y. Chen, J.~Chen, and Y.~Mao, ``A quantum-classical collaborative training architecture based on quantum state fidelity,'' {\em IEEE Transactions on Quantum Engineering}, 2024.

\bibitem{wu2022poster}
J.~Wu and Q.~Li, ``Poster: Scalable quantum convolutional neural networks for edge computing,'' in {\em 2022 IEEE/ACM 7th Symposium on Edge Computing (SEC)}, pp.~307--309, IEEE, 2022.

\bibitem{chen2022quantumLSTM}
S.~Y.-C. Chen, S.~Yoo, and Y.-L.~L. Fang, ``Quantum long short-term memory,'' in {\em ICASSP 2022-2022 IEEE International Conference on Acoustics, Speech and Signal Processing (ICASSP)}, pp.~8622--8626, IEEE, 2022.

\bibitem{li2023pqlm}
S.~S. Li, X.~Zhang, S.~Zhou, H.~Shu, R.~Liang, H.~Liu, and L.~P. Garcia, ``Pqlm-multilingual decentralized portable quantum language model,'' in {\em ICASSP 2023-2023 IEEE International Conference on Acoustics, Speech and Signal Processing (ICASSP)}, pp.~1--5, IEEE, 2023.

\bibitem{yang2022bert}
C.-H.~H. Yang, J.~Qi, S.~Y.-C. Chen, Y.~Tsao, and P.-Y. Chen, ``When bert meets quantum temporal convolution learning for text classification in heterogeneous computing,'' in {\em ICASSP 2022-2022 IEEE International Conference on Acoustics, Speech and Signal Processing (ICASSP)}, pp.~8602--8606, IEEE, 2022.

\bibitem{di2022dawn}
R.~Di~Sipio, J.-H. Huang, S.~Y.-C. Chen, S.~Mangini, and M.~Worring, ``The dawn of quantum natural language processing,'' in {\em ICASSP 2022-2022 IEEE International Conference on Acoustics, Speech and Signal Processing (ICASSP)}, pp.~8612--8616, IEEE, 2022.

\bibitem{stein2023applying}
J.~Stein, I.~Christ, N.~Kraus, M.~B. Mansky, R.~M{\"u}ller, and C.~Linnhoff-Popien, ``Applying qnlp to sentiment analysis in finance,'' in {\em 2023 IEEE International Conference on Quantum Computing and Engineering (QCE)}, vol.~2, pp.~20--25, IEEE, 2023.

\bibitem{stein2021qugan}
S.~A. Stein, B.~Baheri, D.~Chen, Y.~Mao, Q.~Guan, A.~Li, B.~Fang, and S.~Xu, ``Qugan: A quantum state fidelity based generative adversarial network,'' in {\em 2021 IEEE International Conference on Quantum Computing and Engineering (QCE)}, pp.~71--81, IEEE, 2021.

\bibitem{kolle2024quantumDenoisingDiffusionModel}
M.~K{\"o}lle, G.~Stenzel, J.~Stein, S.~Zielinski, B.~Ommer, and C.~Linnhoff-Popien, ``Quantum denoising diffusion models,'' {\em arXiv preprint arXiv:2401.07049}, 2024.

\bibitem{chu2023iqgan}
C.~Chu, G.~Skipper, M.~Swany, and F.~Chen, ``Iqgan: Robust quantum generative adversarial network for image synthesis on nisq devices,'' in {\em ICASSP 2023-2023 IEEE International Conference on Acoustics, Speech and Signal Processing (ICASSP)}, pp.~1--5, IEEE, 2023.

\bibitem{chen2020variational}
S.~Y.-C. Chen, C.-H.~H. Yang, J.~Qi, P.-Y. Chen, X.~Ma, and H.-S. Goan, ``Variational quantum circuits for deep reinforcement learning,'' {\em IEEE access}, vol.~8, pp.~141007--141024, 2020.

\bibitem{jerbi2021parametrized}
S.~Jerbi, C.~Gyurik, S.~Marshall, H.~Briegel, and V.~Dunjko, ``Parametrized quantum policies for reinforcement learning,'' {\em Advances in Neural Information Processing Systems}, vol.~34, pp.~28362--28375, 2021.

\bibitem{skolik2022quantum}
A.~Skolik, S.~Jerbi, and V.~Dunjko, ``Quantum agents in the gym: a variational quantum algorithm for deep q-learning,'' {\em Quantum}, vol.~6, p.~720, 2022.

\bibitem{meyer2022survey}
N.~Meyer, C.~Ufrecht, M.~Periyasamy, D.~D. Scherer, A.~Plinge, and C.~Mutschler, ``A survey on quantum reinforcement learning,'' {\em arXiv preprint arXiv:2211.03464}, 2022.

\bibitem{chen2023efficientQRL_QRC}
S.~Y.-C. Chen, ``Efficient quantum recurrent reinforcement learning via quantum reservoir computing,'' in {\em ICASSP 2024-2024 IEEE International Conference on Acoustics, Speech and Signal Processing (ICASSP)}, pp.~13186--13190, IEEE, 2024.

\bibitem{chen2023quantum_LSTM_RL}
S.~Y.-C. Chen, ``Quantum deep recurrent reinforcement learning,'' in {\em ICASSP 2023-2023 IEEE International Conference on Acoustics, Speech and Signal Processing (ICASSP)}, pp.~1--5, IEEE, 2023.

\bibitem{yun2023quantum}
W.~J. Yun, J.~Park, and J.~Kim, ``Quantum multi-agent meta reinforcement learning,'' in {\em Proceedings of the AAAI Conference on Artificial Intelligence}, vol.~37, pp.~11087--11095, 2023.

\bibitem{chen2024learning}
S.~Y.-C. Chen, ``Learning to program variational quantum circuits with fast weights,'' {\em arXiv preprint arXiv:2402.17760}, 2024.

\bibitem{abbas2021power}
A.~Abbas, D.~Sutter, C.~Zoufal, A.~Lucchi, A.~Figalli, and S.~Woerner, ``The power of quantum neural networks,'' {\em Nature Computational Science}, vol.~1, no.~6, pp.~403--409, 2021.

\bibitem{kuo2021quantum}
E.-J. Kuo, Y.-L.~L. Fang, and S.~Y.-C. Chen, ``Quantum architecture search via deep reinforcement learning,'' {\em arXiv preprint arXiv:2104.07715}, 2021.

\bibitem{ye2021quantum}
E.~Ye and S.~Y.-C. Chen, ``Quantum architecture search via continual reinforcement learning,'' {\em arXiv preprint arXiv:2112.05779}, 2021.

\bibitem{kimura2022quantum}
T.~Kimura, K.~Shiba, C.-C. Chen, M.~Sogabe, K.~Sakamoto, and T.~Sogabe, ``Quantum circuit architectures via quantum observable markov decision process planning,'' {\em Journal of Physics Communications}, vol.~6, no.~7, p.~075006, 2022.

\bibitem{sogabe2022model}
T.~Sogabe, T.~Kimura, C.-C. Chen, K.~Shiba, N.~Kasahara, M.~Sogabe, and K.~Sakamoto, ``Model-free deep recurrent q-network reinforcement learning for quantum circuit architectures design,'' {\em Quantum Reports}, vol.~4, no.~4, pp.~380--389, 2022.

\bibitem{lu2023qas}
X.~Lu, K.~Pan, G.~Yan, J.~Shan, W.~Wu, and J.~Yan, ``Qas-bench: rethinking quantum architecture search and a benchmark,'' in {\em International Conference on Machine Learning}, pp.~22880--22898, PMLR, 2023.

\bibitem{kundu2024enhancing}
A.~Kundu, P.~Bede{\l}ek, M.~Ostaszewski, O.~Danaci, Y.~J. Patel, V.~Dunjko, and J.~A. Miszczak, ``Enhancing variational quantum state diagonalization using reinforcement learning techniques,'' {\em New Journal of Physics}, vol.~26, no.~1, p.~013034, 2024.

\bibitem{sunkel2023ga4qco}
L.~S{\"u}nkel, D.~Martyniuk, D.~Mattern, J.~Jung, and A.~Paschke, ``Ga4qco: genetic algorithm for quantum circuit optimization,'' {\em arXiv preprint arXiv:2302.01303}, 2023.

\bibitem{zhu2023quantum}
X.~Zhu and X.~Hou, ``Quantum architecture search via truly proximal policy optimization,'' {\em Scientific Reports}, vol.~13, no.~1, p.~5157, 2023.

\bibitem{chen2023QRL_QAS}
S.~Y.-C. Chen, ``Quantum reinforcement learning for quantum architecture search,'' in {\em Proceedings of the 2023 International Workshop on Quantum Classical Cooperative}, pp.~17--20, 2023.

\bibitem{selig2023deepqprep}
P.~Selig, N.~Murphy, D.~Redmond, and S.~Caton, ``Deepqprep: Neural network augmented search for quantum state preparation,'' {\em IEEE Access}, 2023.

\bibitem{sun2024quantum}
Y.~Sun, Z.~Wu, Y.~Ma, and V.~Tresp, ``Quantum architecture search with unsupervised representation learning,'' {\em arXiv preprint arXiv:2401.11576}, 2024.

\bibitem{ostaszewski2021reinforcement}
M.~Ostaszewski, L.~M. Trenkwalder, W.~Masarczyk, E.~Scerri, and V.~Dunjko, ``Reinforcement learning for optimization of variational quantum circuit architectures,'' {\em Advances in Neural Information Processing Systems}, vol.~34, pp.~18182--18194, 2021.

\bibitem{wang2023automated}
P.~Wang, M.~Usman, U.~Parampalli, L.~C. Hollenberg, and C.~R. Myers, ``Automated quantum circuit design with nested monte carlo tree search,'' {\em IEEE Transactions on Quantum Engineering}, 2023.

\bibitem{he2023gnn}
Z.~He, X.~Zhang, C.~Chen, Z.~Huang, Y.~Zhou, and H.~Situ, ``A gnn-based predictor for quantum architecture search,'' {\em Quantum Information Processing}, vol.~22, no.~2, p.~128, 2023.

\bibitem{deng2023progressive}
M.~Deng, Z.~He, S.~Zheng, Y.~Zhou, F.~Zhang, and H.~Situ, ``A progressive predictor-based quantum architecture search with active learning,'' {\em The European Physical Journal Plus}, vol.~138, no.~10, p.~905, 2023.

\bibitem{yao2022monte}
J.~Yao, H.~Li, M.~Bukov, L.~Lin, and L.~Ying, ``Monte carlo tree search based hybrid optimization of variational quantum circuits,'' in {\em Mathematical and Scientific Machine Learning}, pp.~49--64, PMLR, 2022.

\bibitem{duong2022quantum}
T.~Duong, S.~T. Truong, M.~Pham, B.~Bach, and J.-K. Rhee, ``Quantum neural architecture search with quantum circuits metric and bayesian optimization,'' in {\em ICML 2022 2nd AI for Science Workshop}, 2022.

\bibitem{wu2023quantumdarts}
W.~Wu, G.~Yan, X.~Lu, K.~Pan, and J.~Yan, ``Quantumdarts: differentiable quantum architecture search for variational quantum algorithms,'' in {\em International Conference on Machine Learning}, pp.~37745--37764, PMLR, 2023.

\bibitem{zhang2022differentiable}
S.-X. Zhang, C.-Y. Hsieh, S.~Zhang, and H.~Yao, ``Differentiable quantum architecture search,'' {\em Quantum Science and Technology}, vol.~7, no.~4, p.~045023, 2022.

\bibitem{sun2024differentiable}
Y.~Sun, J.~Liu, Y.~Ma, and V.~Tresp, ``Differentiable quantum architecture search for job shop scheduling problem,'' {\em arXiv preprint arXiv:2401.01158}, 2024.

\bibitem{liu2023reinforcement}
C.-Y. Liu and H.-S. Goan, ``Reinforcement learning quantum local search,'' in {\em 2023 IEEE International Conference on Quantum Computing and Engineering (QCE)}, vol.~2, pp.~246--247, IEEE, 2023.

\bibitem{fosel2021quantum}
T.~F{\"o}sel, M.~Y. Niu, F.~Marquardt, and L.~Li, ``Quantum circuit optimization with deep reinforcement learning,'' {\em arXiv preprint arXiv:2103.07585}, 2021.

\bibitem{he2022quantum}
Z.~He, C.~Chen, L.~Li, S.~Zheng, and H.~Situ, ``Quantum architecture search with meta-learning,'' {\em Advanced Quantum Technologies}, vol.~5, no.~8, p.~2100134, 2022.

\bibitem{he2022search}
Z.~He, J.~Su, C.~Chen, M.~Pan, and H.~Situ, ``Search space pruning for quantum architecture search,'' {\em The European Physical Journal Plus}, vol.~137, no.~4, p.~491, 2022.

\bibitem{chen2022efficient}
Q.~Chen, Y.~Du, Q.~Zhao, Y.~Jiao, X.~Lu, and X.~Wu, ``Efficient and practical quantum compiler towards multi-qubit systems with deep reinforcement learning,'' {\em arXiv preprint arXiv:2204.06904}, 2022.

\bibitem{ding2022evolutionary}
L.~Ding and L.~Spector, ``Evolutionary quantum architecture search for parametrized quantum circuits,'' in {\em Proceedings of the Genetic and Evolutionary Computation Conference Companion}, pp.~2190--2195, 2022.

\bibitem{zhang2023evolutionary}
A.~Zhang and S.~Zhao, ``Evolutionary-based searching method for quantum circuit architecture,'' {\em Quantum Information Processing}, vol.~22, no.~7, p.~283, 2023.

\bibitem{ding2023multi}
L.~Ding and L.~Spector, ``Multi-objective evolutionary architecture search for parameterized quantum circuits,'' {\em Entropy}, vol.~25, no.~1, p.~93, 2023.

\bibitem{subasi2023toward}
O.~Subasi, ``Toward automated quantum variational machine learning,'' {\em arXiv preprint arXiv:2312.01567}, 2023.

\bibitem{sun2023differentiable}
Y.~Sun, Y.~Ma, and V.~Tresp, ``Differentiable quantum architecture search for quantum reinforcement learning,'' in {\em 2023 IEEE International Conference on Quantum Computing and Engineering (QCE)}, vol.~2, pp.~15--19, IEEE, 2023.

\bibitem{zhang2021neural}
S.-X. Zhang, C.-Y. Hsieh, S.~Zhang, and H.~Yao, ``Neural predictor based quantum architecture search,'' {\em Machine Learning: Science and Technology}, vol.~2, no.~4, p.~045027, 2021.

\bibitem{du2022quantum}
Y.~Du, T.~Huang, S.~You, M.-H. Hsieh, and D.~Tao, ``Quantum circuit architecture search for variational quantum algorithms,'' {\em npj Quantum Information}, vol.~8, no.~1, p.~62, 2022.

\bibitem{schuld2019evaluating}
M.~Schuld, V.~Bergholm, C.~Gogolin, J.~Izaac, and N.~Killoran, ``Evaluating analytic gradients on quantum hardware,'' {\em Physical Review A}, vol.~99, no.~3, p.~032331, 2019.

\bibitem{bergholm2018pennylane}
V.~Bergholm, J.~Izaac, M.~Schuld, C.~Gogolin, C.~Blank, K.~McKiernan, and N.~Killoran, ``Pennylane: Automatic differentiation of hybrid quantum-classical computations,'' {\em arXiv preprint arXiv:1811.04968}, 2018.

\bibitem{chen2022variational}
S.~Y.-C. Chen, C.-M. Huang, C.-W. Hsing, H.-S. Goan, and Y.-J. Kao, ``Variational quantum reinforcement learning via evolutionary optimization,'' {\em Machine Learning: Science and Technology}, vol.~3, no.~1, p.~015025, 2022.

\bibitem{caro2022generalization}
M.~C. Caro, H.-Y. Huang, K.~Sharma, A.~Sornborger, L.~Cincio, and P.~J. Coles, ``Generalization in quantum machine learning from few training data,'' {\em Nature communications}, vol.~13, no.~1, p.~4919, 2022.

\bibitem{sutton2018reinforcement}
R.~S. Sutton and A.~G. Barto, {\em Reinforcement learning: An introduction}.
\newblock MIT press, 2018.

\bibitem{mnih2015human}
V.~Mnih, K.~Kavukcuoglu, D.~Silver, A.~A. Rusu, J.~Veness, M.~G. Bellemare, A.~Graves, M.~Riedmiller, A.~K. Fidjeland, G.~Ostrovski, {\em et~al.}, ``Human-level control through deep reinforcement learning,'' {\em nature}, vol.~518, no.~7540, pp.~529--533, 2015.

\bibitem{doubledqn}
H.~Van~Hasselt, A.~Guez, and D.~Silver, ``Deep reinforcement learning with double q-learning,'' in {\em Proceedings of the AAAI conference on artificial intelligence}, vol.~30, 2016.

\bibitem{zhang2022tensorcircuit}
S.-X. Zhang, J.~Allcock, Z.-Q. Wan, S.~Liu, J.~Sun, H.~Yu, X.-H. Yang, J.~Qiu, Z.~Ye, Y.-Q. Chen, {\em et~al.}, ``Tensorcircuit: a quantum software framework for the nisq era,'' {\em arXiv preprint arXiv:2205.10091}, 2022.

\bibitem{paszke2019pytorch}
A.~Paszke, S.~Gross, F.~Massa, A.~Lerer, J.~Bradbury, G.~Chanan, T.~Killeen, Z.~Lin, N.~Gimelshein, L.~Antiga, {\em et~al.}, ``Pytorch: An imperative style, high-performance deep learning library,'' {\em Advances in neural information processing systems}, vol.~32, 2019.

\bibitem{scikit-learn}
F.~Pedregosa, G.~Varoquaux, A.~Gramfort, V.~Michel, B.~Thirion, O.~Grisel, M.~Blondel, P.~Prettenhofer, R.~Weiss, V.~Dubourg, J.~Vanderplas, A.~Passos, D.~Cournapeau, M.~Brucher, M.~Perrot, and E.~Duchesnay, ``Scikit-learn: Machine learning in {P}ython,'' {\em Journal of Machine Learning Research}, vol.~12, pp.~2825--2830, 2011.

\end{thebibliography}

\end{document}